\documentclass{article}[12pt]
\usepackage{geometry} 
\usepackage{setspace}
\usepackage[T1]{fontenc}
\usepackage{titlesec}
\usepackage{amsmath, amssymb, amsfonts, amsthm}
\usepackage{bm}
\usepackage{graphicx}
\usepackage[title,titletoc]{appendix}
\usepackage{apptools}
\usepackage{color}
\usepackage{multirow}
\usepackage{setspace}
\usepackage{hyperref}
\usepackage{float}
\usepackage{indentfirst}
\usepackage{enumitem}
\usepackage{xfrac}
\usepackage{mathtools}
\usepackage{caption}
\usepackage{subcaption}
\usepackage{tabularx}
\usepackage{booktabs}
\usepackage{siunitx}
\usepackage[authoryear]{natbib}
\usepackage{pdflscape}
\usepackage{listings}
\usepackage{xcolor}
\usepackage{pgfplots}
\usepackage{makecell}
\usepackage[super]{nth}
\usepackage{xr}
\usepackage{bigdelim}  
\usepackage{pifont}
\usepackage{nccmath,amsmath,lipsum}
\usepackage{etoolbox}
\usepackage{appendix}
\usepackage{setspace}
\BeforeBeginEnvironment{appendices}{\clearpage}

\setlist[itemize]{noitemsep,nolistsep}
\setlist[enumerate]{noitemsep,nolistsep}
\numberwithin{figure}{section}
\numberwithin{table}{section}
\numberwithin{equation}{section} 
\makeatletter
\@addtoreset{equation}{section}
\makeatother

\graphicspath{{./Pictures/}{./Chapter1}}

\title{Imputation of Nonignorable Missing Data in Surveys Using Auxiliary Margins Via Hot Deck and Sequential Imputation}

\author{Yanjiao Yang\thanks{Department of Statistical Science, Box 90251, Duke University, Durham, NC 27708-0251} \and Jerome P. Reiter\thanks{Department of Statistical Science, 214a Old Chemistry Building, Duke University, Durham, NC 27708-0251 }}
\date{}

\begin{document}

\maketitle

\begin{abstract}
Survey data collection often is plagued by unit and item nonresponse. To reduce reliance on strong assumptions about the missingness mechanisms,  statisticians can use information about population marginal distributions known, for example, from censuses or administrative databases. One approach that does so is the Missing Data with Auxiliary Margins, or MD-AM, framework, which  uses multiple imputation for both unit and item nonresponse so that survey-weighted estimates accord with the known marginal distributions.  However, this framework relies on specifying and estimating a joint distribution for the survey data and nonresponse indicators, which can be computationally and practically daunting in data with many variables of mixed types.  We propose two adaptations to the MD-AM framework to simplify the imputation task.  First, rather than specifying a joint model for unit respondents' data, we use random hot deck imputation while still leveraging the known marginal distributions.  Second, instead of sampling from conditional distributions implied by the joint model for the missing data due to item nonresponse, we apply multiple imputation by chained equations for item nonresponse before imputation for unit nonresponse. Using simulation studies with nonignorable missingness mechanisms, we demonstrate that the proposed approach can provide more accurate point and interval estimates than models that do not leverage the auxiliary information. We illustrate the approach using data on voter turnout from the U.S.\ Current Population Survey. \\

\end{abstract}

\section{Introduction}\label{sec:intro}

When data suffer from nonresponse, survey organizations typically must make strong and unverifiable assumptions about the reasons for missingness. For example, they may have to assume that values are missing at random  \citep{rubin1976inference}.  When unrealistic, such assumptions may result in inaccurate estimates about population quantities or inferences about model parameters. 

To reduce reliance on assumptions, survey organizations may be able to take advantage of information on marginal distributions available in external data sources.  For example, they may know population totals or percentages of demographic variables from censuses, administrative databases, or high quality surveys.  Indeed, this is a common situation in practice, as survey organizations often use calibration, post-stratification, or raking techniques to ensure survey-weighted estimates  align with known population quantities \citep[e.g., ][]{deville1992calibration, nevo2003using, lumley2011complex}.  Much of this work relies on adjusting survey participants' weights to account for unit nonresponse and some form of imputation for item nonresponse.  

We consider an alternative paradigm in which multiple imputation \citep{rubin1987multiple} is used for all missing values due to nonresponse. This can have some potential advantages relative to the re-weighting plus imputation approach \citep{peytchev:poq, little:discussion, alanya:etal, dever:valliant}, as we discuss in Section \ref{sec:conc}. In particular, we build on the Missing Data with Auxiliary Margins, or MD-AM, framework proposed by \citet{akande2021leveraging}.  They incorporate known marginal distributions in multiple imputation for categorical variables in simple random samples. They use a selection modeling approach \citep{little:rubin}, i.e.,  a joint model for the survey variables coupled with models for the unit and item nonresponse indicators given the survey variables, that allows for potentially different missingness mechanisms for unit and item nonresponse, e.g., missing at random for item nonresponse and  missing not at random for unit nonresponse. \citet{akande2021multiple} further develop the MD-AM framework for stratified simple random samples.  Their key innovation is an imputation procedure  that results in plausible completed-data, survey-weighted estimates considering the known margins.  \citet{tang2024using} introduce an MD-AM model that relies on selection models for the item nonresponse indicators and a pattern mixture model for the unit nonresponse indicator.  This hybrid missingness MD-AM model enables analysts to leverage auxiliary marginals in the multiple imputation while accounting generally for unequal probabilities of selection.

The hybrid missingness MD-AM model of \citet{tang2024using} has some implementation challenges.  First, as a parametric model-based framework, it requires formulating a joint distribution for the data values and the missingness indicators.  This can be a daunting task when many variables are subject to nonresponse.  Inferences may be sensitive to the model specification, particularly when rates of unit nonresponse are high, since all unit nonrespondents' values are imputed.  Second, even with an accurate specification, their modeling strategy can require computationally intensive Monte Carlo steps.  Third, and related to the previous two challenges, their strategy relies on parametric modeling, whereas increasingly multiple imputation is done using algorithmic techniques like classification and regression trees \citep[CART, ][]{burgette:reiter}. Finally, it does not consider continuous variables, which are sometimes present in surveys, e.g., yearly age.

Motivated by these challenges, we build on the hybrid missingness MD-AM framework with the aim of developing imputation approaches that flexibly can handle multivariate data. To do so, we partly replace the parametric modeling for unit nonresponse with hot deck imputation \citep{mander2003weighted, andridge2009use}, 
with the goal of gaining some immunity to model misspecification. As a further enhancement, we replace the computationally intensive Monte Carlo steps for item nonresponse imputation with an implementation of multiple imputation by chained equations \citep{raghunathan2001multivariate, vanbuuren}.  This allows analysts to separate the imputation process into two steps, including item nonresponse imputation that utilizes existing software. With these two innovations, the result is a multiple imputation approach that uses auxiliary margins, accounts for survey weights, and  allows for convenient application. 

The remainder of this article is organized as follows.
In Section \ref{sec:background}, we review the hybrid missingness MD-AM modeling framework as presented in \citet{tang2024using}. In Section \ref{sec:framework},  we describe how we integrate hot deck imputation for unit nonresponse into the MD-AM modeling framework, as well as the two-stage approach that utilizes multiple imputation by chained equations for item nonresponse. In Section \ref{sec:sims}, we present simulation studies that illustrate the repeated sampling properties of the proposed approach. In Section \ref{sec:CPS}, we present an example application of the approach using data on voter turnout from the U.S.\ Current Population Survey (CPS).  Finally, in Section \ref{sec:conc}, we conclude with directions for future research. Code for the simulation studies and CPS analysis can be found at \url{https://github.com/yjyang00/MNAR}.

\section{Review: Hybrid Missingness MD-AM Model with Survey Weights}\label{sec:background}

Suppose we have a finite population comprising $N$ individuals having $k$ survey variables, $\boldsymbol{X}=(X_1, \dots, X_k)$.  For $i=1, \dots, N$, let $\boldsymbol{x}_i = (x_{i1}, \dots, x_{ik})$ be the survey variables for individual $i$.  
For $i=1, \dots, N$, let $I_{i}=1$ when individual $i$ is selected to be in the survey, and let $I_i=0$ otherwise; let $\pi_i = Pr(I_i=1)$ be the probability that individual $i$ is selected to be in the survey; and, let $w^d_i = 1/\pi_i$ be the design weight for individual $i$.  Let $n=\sum_{i=1}^N I_i$ be the intended sample size in the survey, which we refer to as $\mathcal{S}$.

We suppose that $\mathcal{S}$ suffers from unit and item nonresponse. For all units sampled in $\mathcal{S}$, we define the unit nonresponse indicator $U$ so that $U_i=1$ when sampled unit $i$ does not provide answers to any questions in the survey and $U_i=0$ otherwise. We also define item nonresponse indicators $\boldsymbol{R}=(R_1, \dots, R_k)$ corresponding to $(X_1, \dots, X_k)$ so that, for any unit $i$ that participates in the survey, i.e., $U_i=0$, and any variable $X_j$, we have  $R_{ij}=1$ if unit $i$ does not respond to the question concerning $X_j$ and $R_{ij}=0$ otherwise.  When $U_i=1$, $(R_{i1}, \dots, R_{ik})$ is not defined.  

We presume access to some set of auxiliary information  $\mathcal{A}$ on the marginal distributions of $\boldsymbol{X}$.  Here, we assume $\mathcal{A}$ comprises accurate estimates of univariate population percentages or totals of some subset of the categorical variables in $\boldsymbol{X}$. We write $X_j \in \mathcal{A}$ to indicate that $X_j$ has auxiliary margins in $\mathcal{A}$.  In what follows, we assume the uncertainty associated with these population values is negligible.  In Section \ref{sec:conc} we mention how to adapt the model when this is not the case.   
 
We seek to use multiple imputation for any  $x_{ij} \in \mathcal{S}$ that is not observed, whether arising from unit or item nonresponse, while leveraging the information in $\mathcal{A}$. We also wish to ensure any imputations result in plausible survey-weighted inferences relative to the known marginals.  We do so via the hybrid missingness MD-AM model of \citet{tang2024using}, which we now summarize.

\subsection{General Model Specification}\label{sec:modelnow}
In the hybrid missingness MD-AM framework, we specify a joint distribution for $(\boldsymbol{X}, \boldsymbol{R}, U)$, using the information in $\mathcal{A}$. 
We first specify a model for $U$ and a model for $\boldsymbol{X}|U$, i.e., we rely on a pattern mixture model for unit nonresponse.  We then specify a model for $\boldsymbol{R}|\boldsymbol{X}, U$, i.e., we rely on selection models for item nonresponse.
\citet{tang2024using}  use a sequential factorization of the joint distribution, beginning with the model for $U$.  For all $i$ with $I_i=1$, we assume  
\begin{equation}\label{Umodel}
U_i \sim Bernoulli(\pi_U). 
\end{equation}
We specify $f(\boldsymbol{X}|U) = f(X_1 | U)(f(X_2|X_1, U) \cdots f(X_k|X_1, \dots, X_{k-1},U)$. This requires an ordering of the conditional models.  For now, we assume an ordering with $X_1$ first, $X_2$ second, and so on until $X_k$. \citet{tang2024using} offer advice about setting the ordering, which mainly comes to embedding modeling preferences based on domain knowledge and facilitating computation.
For related discussion, see \citet{si:reiter:hillygus:pa} and \citet{bart:daniels}.
For any $j$, we let $\Omega_j$ and $\theta_j$ represent parameters in the conditional model for $X_j$. When  $X_j$ has a margin,  let $I(X_j \in \mathcal{A})=1$. When  $X_j$ does not have a margin, let $I(X_j \in \mathcal{A})=0$. 

For $X_1$,  we specify the model  
\begin{equation}\label{modelx1}
X_{1} \mid U \sim g_1(\Omega_{1}, \theta_{1} U I(X_1 \in \mathcal{A})).
\end{equation}
Here,  $g_1(-)$ is some model parameterized by an additive function of its operands.  For example, if $X_1$ is binary then $g_1$ could be a logistic regression with an intercept  
$\Omega_1=\omega_{10}$ and, if $X_1$ has a margin in $\mathcal{A}$, a main effect for $U$ with coefficient $\theta_1$. Thus, when $\theta_1\neq 0$, \eqref{modelx1} implies a different distribution of $X_1$ for unit respondents and nonrespondents.  For $X_2$, we specify   
\begin{equation}
X_{2} \mid X_{1}, U \sim g_2(X_{1}, \Omega_{2}, \theta_2 U I(X_2 \in \mathcal{A})) 
\end{equation}
where $g_2(-)$ is some regression whose linear predictor has 
an intercept $\omega_{20}$, a main effect for $X_1$ with coefficient $\omega_{21}$, and, if $X_2$ has a margin in $\mathcal{A}$, a main effect for $U$ with coefficient $\theta_2$.  This sequence continues until $X_k$, where for any $X_j$ we have  
\begin{equation}
X_{j} \mid X_{1}, \dots, X_{j-1}, U \sim g_j(X_{1}, \dots, X_{j-1}, \Omega_j, \theta_j U I(X_j \in \mathcal{A})),
\end{equation}
where $g_j(-)$ is some function of $X_1, \dots, X_{j-1}$, including possibly some interaction terms, that is linear in the coefficients in $\Omega_j$ and has a main effect for $U$ if $X_j \in \mathcal{A}$. 

We specify the models for $(R_1, \dots, R_k)$ in the same order as the models for $(X_1, \dots, X_k)$.  For any arbitrary $R_j$, \citet{tang2024using} assume that  
\begin{equation}\label{Rmodel}
R_{j} \mid X_{1}, \dots, X_{k}, U=0 \sim Bernoulli(\pi_{R_{j}}),\,\,\, logit(\pi_{R_{j}}) =  h_j(X_{1}, \dots, X_{j-1}, X_{j+1}, \dots X_{k}, \Phi_j). 
\end{equation}
Each $h_j(-)$ is some predictor function that does not include its corresponding $X_j$ but can include main effects and interaction effects involving the other variables in $\boldsymbol{X}$. \citet{tang2024using} do not include other item nonresponse indicators as predictors in \eqref{Rmodel} for modeling convenience.  

The hybrid missingness MD-AM model makes two key identifying assumptions.  First, the pattern mixture  models for $\boldsymbol{X}|U$ do not include interactions involving $U$.  This is because the univariate marginals in  $\mathcal{A}$ only allow for identification of main effects of $U$ \citep{si:reiter:hillygus:aoas, sadinle2019sequentially}.  Interactions involving $U$ can be identified when $\mathcal{A}$ includes bivariate marginals.  Second, the model assumes that item nonresponse follows an itemwise conditionally independent nonresponse (ICIN) mechanism \citep{sadinle2017itemwise}.  In this mechanism, $R_j$ is conditionally independent of $X_j$ but possibly dependent on other variables. We assume ICIN because $\mathcal{A}$ does not provide enough information to allow both $U$ and $R_j$ to depend on $X_j$.  

\subsection{Incorporating Survey Weights}
\label{weight_UNa}

We now review how \citet{tang2024using} incorporate survey weights into the hybrid missingness MD-AM model. For any categorical $X_j$ taking on levels $\{1, \dots, m_j\}$ and having a margin in $\mathcal{A}$, let $T_{X^c_j} = \sum_{i=1}^NI(x_{ij}=c)$ be the total number of individuals in the population at level $c \in \{1, \dots, m_j\}$ of $X_j$.  We work with population totals, although one also could work with population percentages. 
With complete data, we would estimate  $T_{X^c_j}$ with the \citet{horvitz1952generalization} estimator, 
$\hat{T}_{X^c_j} = \sum_{i \in \mathcal{S}}w_i^d I(x_{ij}=c)$.
When $n$ is large enough and data are fully observed,   
finite population central limit theorems assure that $\hat{T}_{X^c_j}$ follows a normal distribution centered at $T_{X^c_j}$ with some variance that can be estimated, for example, using design-based principles. When considering the estimated totals for all levels of $X_j$ simultaneously, they approximately follow a multivariate normal distribution centered at the mean vector $(T_{X^1_j}, \dots, T_{X^{m_j}_j})$ with some estimable covariance. In what follows, for convenience we use the marginal sampling distributions for the total at each level.

\citet{akande2021multiple} and \citet{tang2024using} suggest that any imputations of the missing values for $X_j$ with a margin in $\mathcal{A}$ result in a completed data set with  plausible values of $\hat{T}_{X^c_j}$ relative to the known ${T}_{X^c_j}$. 
In particular, for all $i \in \mathcal{S}$ and for all $j$, let $x_{ij}^\star = x_{ij}$ when $R_{ij} = 0$, and let $x_{ij}^\star$ be an imputed value when $R_{ij} = 1$ or $U_i=1$. In addition to  \eqref{Umodel} -- \eqref{Rmodel}, the imputations of missing values of $X_j$ when $X_j \in \mathcal{A}$ should satisfy, 
\begin{equation}
\sum_{i \in \mathcal{S}}w_i^d I(x_{ij}^{\star} = c) \sim N(T_{X^c_j},V_{X^c_j}).
\label{prob_cons}
\end{equation}
\citet{tang2024using} describe $V_{X^c_j}$ as a ``parameter set by the analyst to reflect how closely [$\hat{T}_{X^c_j}$] should match [$T_{X^c_j}$] in any completed data set.''  A small $V_{X^c_j}$  encourages imputations that make $\hat{T}_{X^c_j}$ close to $T_{X^c_j}$. As a default, they suggest setting $V_{X^c_j}$ to approximate the sampling variance of $\hat{T}_{X^c_j}$ had there been no missing data in $X_j$. We discuss methods for doing so in the context of the simulations in Section \ref{sec:sims} and CPS data analysis in Section \ref{sec:CPS}. 

The procedure for estimating the hybrid missingness MD-AM model requires weights for all units. \citet{tang2024using} presume that the design weights for unit nonrespondents are not available, as is typical in released data files in practice, in which case we need to create them. \cite{tang2024using} suggest using weights
\begin{equation}
w_i =  \left\{\begin{array}{lr}
        w_i^d & \text{if } U_i = 0\\
        \frac{N-\sum_{j \in \mathcal{S}} w_j^d}{\sum_{j \in \mathcal{S}} U_j} & \text{if } U_i = 1.
        \end{array}\right. \label{known_weights_unit_res}
\end{equation}
 \citet{tang2024using} also describe a procedure for approximating design weights when these are not available, for example, because an agency provides adjusted weights due to calibration of post-stratification on the released survey file.  We present this procedure in the context of the CPS data analysis in Section \ref{sec:CPS}.

\subsection{Estimation Strategy}\label{sec:estimation}

To estimate the model and impute missing values, \citet{tang2024using} propose an iterative sampler. 
The algorithm cycles between drawing model parameters given the completed data, imputing missing survey variables due to item response, and imputing all survey variables for the unit nonrespondents. Here we outline the estimation approach in broad terms.  

We sample draws of parameters not attached to $U$ terms in the models for $\boldsymbol{X}|U$, i.e., the $\Omega_j$ parameters, by using the completed data for unit respondents. To ease computations, we use normal approximations for the distributions of these parameters.  Similarly, we sample the parameters for item nonresponse, i.e., the $\Phi_j$ parameters, also using normal approximations and the completed data for unit respondents.  We describe the sampling steps for the $\theta_j$ parameters attached to $U$ terms when we discuss the imputation of $\boldsymbol{X}$ for unit nonrespondents.

Given the draws of $(\Omega_j, \Phi_j)$, we sample a value for any missing $x_{ij}$ due to item nonresponse from its conditional distribution.  Here, we condition on the current draw of the completed-data other than $x_{ij}$, which we write for unit $i$ as $\boldsymbol{x}_{i(-j)}^*$. As a general expression, we have 
\begin{equation}\label{eq:imputeitem}
p(x_{ij} | \boldsymbol{R}, U_i=0, \Omega_j, \boldsymbol{x}_{i(-j)}^*) \propto \Pi_{t \neq j}Pr(R_{it}=1 | U_i=0, \Omega_j, \boldsymbol{x}_{i}^*) p(\boldsymbol{x}_{i}^* | U_i=0, \Omega_j, \boldsymbol{x}_{i(-j)}^*).
\end{equation}
We provide specific examples of \eqref{eq:imputeitem} in Section \ref{sec:sims}.  With large $k$, \eqref{eq:imputeitem} is a product of many distributions, which can slow down running time. Further, when some $X_j$ are continuous---a case not covered by \citet{tang2024using}---sampling from \eqref{eq:imputeitem}  can involve rejection sampling steps. The coding and computational challenges inherent with such steps partly motivate our approach, as we describe in Section \ref{sec:framework} and Section \ref{sec:sims}.

To sample values of $\boldsymbol{X}$ for units with $U_i=1$, we use the intercept matching algorithm described in \citet{tang2024using}.  The key step in the  algorithm is to obtain a draw of any $\theta_j$ parameters in the models for $\mathbf{X}|U$. For each $X_j \in \mathcal{A}$ and each level $c$ of that variable, we sample a plausible value of the complete-data Horvitz-Thompson estimator, say $\hat{T}_{X^c_j}$.  For convenience, we do so by sampling values marginally from $\hat{T}_{X^c_j} \sim N(T_{X^c_j}, V_{X^c_j})$ for $m_j-1$ of the levels of $X_j$; refer to the set of selected levels as $C$. We set the value of $\hat{T}_{X^c_j}$ for the remaining level to $N - \sum_{c \in C} \hat{T}_{X^c_j}$, using a survey-weighted estimate of $N$ if the population size is not known exactly.  Alternatively, we could sample plausible values of $(\hat{T}_{X^1_j}, \dots, \hat{T}_{X^{m_j}_j})$ following a multivariate normal modeling approach as in \cite{schifeling:reiter:fusion}. 
With either approach, we then determine the distribution of imputed values that would make the Horvitz-Thompson estimates in the completed data approximately equal to $(\hat{T}_{X^1_j}, \dots, \hat{T}_{X^{m_j}_j})$; call this distribution $(p_j^1, \dots, p_j^{m_j})$.  We set each $\theta_j$ so that predicted probabilities in the model for $X_j|U=1$ match $(p_j^1, \dots, p_j^{m_j})$ on average. Using these  $\theta_j$ parameters and the current draw of each $\Omega_j$, we impute the missing $x_{ij}$ for the unit nonrespondents for all $X_j \in \mathcal{A}$. For more details on this process, see Appendix \ref{appendixA} which describes the steps of the intercept matching algorithm for the CPS data analysis. 
 For the remaining $X_j$ without margins, we impute the missing $x_{ij}$ for the unit nonrespondents from the models for $X_j|X_1, \dots, X_{j-1}, U=0$.  

\section{Hot Deck and MICE in Hybrid Missingness MD-AM Models} 
\label{sec:framework}

While the hybrid missingness MD-AM model leverages auxiliary margins and accounts for survey weights in imputations, as noted previously, implementation of the model presents several practical challenges. We therefore introduce two innovations, namely (i) replacing parts of the model-based imputations for unit nonrespondents with hot deck imputations as described in Section \ref{sec:hotdeck}, and (ii) replacing the sampling steps in \eqref{eq:imputeitem} with an application of multiple imputation by chained equations as described in Section \ref{sec:mice}. 

\subsection{Adding the Hot Deck for Unit Nonresponse}\label{sec:hotdeck}

In hot deck imputation, analysts match units with missing values (known as recipients) to units with observed values (known as donors) based on their  similarity on some set of variables. The donor's values are used as the imputation for the recipient's missing values. There are various types of hot deck imputation methods  \citep{kalton1986treatment, andridge2010review}. We use the random hot deck, in which each unit's donor is randomly selected from a pool of possible donors. Other methods include, for example,  nearest neighbor hot deck, weighted sequential hot deck, hierarchical hot deck, and fractional hot deck \citep{cox1980weighted, brick1996handling, kim2004fractional}. 

For each unit nonrespondent, we first impute its  values for each $X_j \in \mathcal{A}$ following the intercept matching algorithm outlined in Section \ref{sec:estimation}. Then, we employ a random hot deck to impute the remaining survey variables without known  margins. To create the donor pools, we form cross-classifications of all combinations of the survey variables with known margins. For each unit with $U_i=1$, we randomly select a donor from its matching pool. For instance, suppose $X_1$ and $X_2$ are binary variables with known margins, and $X_3$ and $X_4$ are variables without margins. We form four donor pools from all units in $\mathcal{S}$ with  $U_i=0$, each defined as the set of all units with the same values of $(x^*_{i1}, x^*_{i2}) \in \{(0,0), (0,1), (1,0), (1,1)\}$. For each unit nonrespondent, we identify the donor pool corresponding to their imputed values for $(X_1, X_2)$ and randomly sample a value of $(X_3, X_4)$ from the donors in that pool. For example, if the imputation of $(X_1, X_2)$ for unit $i$ is $(x_{i1}^*, x_{i2}^*) =(0,0)$, we sample $(x_{i3}^*, x_{i4}^*)$ from the records in the $(0,0)$ donor pool. 

\subsection{Adding MICE for Item Nonresponse}\label{sec:mice}
The hot deck simplifies imputation for the unit nonrespondents, but it does not resolve the computational challenges for item nonresponse imputation.  To do so, we break the imputations into two distinct stages.  Instead of using \eqref{eq:imputeitem} inside an iterative estimation algorithm, we implement a multiple imputation by chained equations (MICE) algorithm using the records with $U_i=0$ to create $L$ completed datasets for the survey participants.  Once we have each set of completed data for item nonresponse, we use the intercept matching algorithm with hot deck imputation to generate the imputations for unit nonrespondents, resulting in a completed data set for all $n$ records in $\mathcal{S}$.  The MICE algorithm can use nonparametric imputation engines, such as CART models or predictive mean matching. Even with parametric models, using MICE instead of \eqref{eq:imputeitem} has the appealing feature of not requiring rejection sampling steps when imputing missing values of continuous variables.

\section{Simulation Studies}\label{sec:sims}
As a finite population, we generate data for $N=3,373,378$ units, which is the number of individuals in the 2022 American Community Survey (ACS) data file downloaded from the U.S.\ Census Bureau's Microdata Access Tool website (\url{https://data.census.gov/}). The file includes a set of weights, which we turn into a size variable $Z$ that is fully observed as $(z_1, \dots, z_N)$.  We use $Z$ to define the first-order inclusion probabilities for Poisson sampling, setting each $\pi_i=1/10z_i$. In this way, we expect a sample size of approximately $n=6000$ units in any Poisson sample.  Let $W=10Z$ be the survey weights. 

For each unit $i=1, \dots, N$, we generate four binary variables $(x_{i1}, x_{i2}, x_{i3}, x_{i4})$ and two continuous variables $(x_{i5}, x_{i6})$.  We allow $X_1$ to depend on $W$ so that the sampled data include associations between the survey weights and the survey variables. The data generation model is displayed in \eqref{2a}--\eqref{2g}. 
\begin{align}
\label{2a}
U &\sim \text{Bernoulli}(\pi_{U})\quad \text{logit}(\pi_{U})=\nu_0\\
\label{2b}
X_1|U &\sim \text{Bernoulli}(\pi_{x_1})\quad 
\text{logit}(\pi_{x_1})=\omega_{10}+\omega_{11}W + \theta_{1}U \\
\label{2c}
X_2|X_1,U&\sim \text{Bernoulli}(\pi_{x_2})\quad
\text{logit}(\pi_{x_2})=\omega_{20}+\omega_{21}X_1+\theta_{2}U\\
\label{2d}
X_3|X_1,X_2,U&\sim \text{Bernoulli}(\pi_{x_3})\quad
\text{logit}(\pi_{x_3})=\omega_{30}+\omega_{31}X_1+\omega_{32}X_2
\\
\label{2e}
 X_4|X_1,X_2,X_3,U&\sim \text{Bernoulli}(\pi_{x_4})\quad
\text{logit}(\pi_{x_4})=\omega_{40}+\omega_{41}X_1+\omega_{42}X_2+\omega_{43}X_3
\\
\label{2f}
X_5|X_1,X_2,X_3,X_4,U &\sim \text{Normal}(\omega_{50}+\omega_{51}X_1+\omega_{52}X_2+\omega_{53}X_3+\omega_{54}X_4,\sigma_{x_5}^2)\\
\label{2g}
X_6|X_1,X_2,X_3,X_4,X_5,U &\sim \text{Normal}(\omega_{60}+\omega_{61}X_1+\omega_{62}X_2+\omega_{63}X_3+\omega_{64}X_4+\omega_{65}X_5,\sigma_{x_6}^2).
\end{align}

We set $\nu_0=-1.2$ so that the unit nonresponse rate is around 23\%.   We let $\theta_{1}\in\{-2,-0.5\}$, where $\theta_1=-2$ generates a substantial difference in the marginal distribution of $X_1$ for unit respondents and nonrespondents whereas $\theta_1=-0.5$ corresponds to a more modest difference.  We set $\theta_2 = -2$ to correspond to strongly nonignorable unit nonresponse related to $X_2$. We set other parameters so that there are associations among the variables. Specific parameter values are available in Appendix \ref{appendixB}.

We independently sample 500 datasets using Poisson sampling. For each sample $\mathcal{S}$, we introduce sample-specific unit nonresponse by replacing the $U_i$ generated in \eqref{2a} with a random draw from the distribution of $U|X_1, X_2$.  To do so, we compute $P(U=1|X_1, X_2)$ for each of the four combinations of $(X_1, X_2)$ using the true probabilities determined by joint distribution of $(U, X_1, X_2)$ per \eqref{2a}--\eqref{2c}.  We then take a Bernoulli draw of each $U_i$ based on $(x_{i1}, x_{i2})$, and we set $\boldsymbol{x}_i$ as completely missing for any unit with a randomly sampled $U_i=1$.  For item nonresponse, we randomly sample missingness indicators according to ICIN mechanisms.
\begin{align}
\label{2h}
R_2|X_1, X_2, X_3, X_4, X_5, X_6, U=0
&\sim \text{Bernoulli}(\pi_{R_2})\\\nonumber
\text{logit}(\pi_{R_{2}})&=\phi_{20}+\phi_{21}X_1+\phi_{22}X_3+\phi_{23}X_4+\phi_{24}X_5+\phi_{25} X_6
\\
\label{2i}
R_3|X_1, X_2, X_3, X_4, X_5, X_6, U=0
&\sim \text{Bernoulli}(\pi_{R_{3}})\\\nonumber
\text{logit}(\pi_{R_{3}})&=\phi_{30}+\phi_{31}X_1+\phi_{32}X_2+\phi_{33}X_4+\phi_{34}X_5+\phi_{35}X_6\\
\label{2j}
R_4|X_1, X_2, X_3, X_4, X_5, X_6, U=0
&\sim \text{Bernoulli}(\pi_{R_{4}})\\\nonumber
\text{logit}(\pi_{R_{4}})&=\phi_{40}+\phi_{41}X_1+\phi_{42}X_2+\phi_{43}X_3+\phi_{44}X_5+\phi_{45}X_6
\\
\label{2k}
R_6|X_1, X_2, X_3, X_4, X_5, X_6, U=0
&\sim \text{Bernoulli}(\pi_{R_{6}})\\\nonumber
\text{logit}(\pi_{R_{6}})&=\phi_{60}+\phi_{61}X_1+\phi_{62}X_2+\phi_{63}X_3+\phi_{64}X_4+\phi_{65}X_5.
\end{align}
We delete any value in $\mathcal{S}$ for which the sampled $R_{ij}=1$. We set the item nonresponse parameters to generate approximately 25\% item nonresponse in each eligible survey variable. Specific parameter values are available in Appendix \ref{appendixB}. We do not introduce any item nonresponse to $X_1$ and $X_5$, mainly to illustrate scenarios with some complete variables.  

After making the missing data in each $\mathcal{S}$, we estimate three hybrid missingness MD-AM models.  The first implements both innovations from Section \ref{sec:framework}. It uses a bespoke implementation of the ``mice'' package in R \citep{vanbuuren} to generate completed data for item nonresponse and hot deck imputation with the intercept matching algorithm to generate the completed data for unit nonrespondents.  We refer to this method as ``MMH'' for the MD-AM model with MICE and hot deck. The second implements only the innovation in Section \ref{sec:hotdeck}. It follows the iterative estimation algorithm  in \citet{tang2024using} to generate imputations for item nonrespondents and uses the hot deck imputation with the intercept matching algorithm for unit nonrespondents.  We refer to this ``MH'' for the MD-AM model with hot deck. We also fit a fully parametric version of the hybrid missingness MD-AM model using the approach in \citet{tang2024using}.  We refer to this as the ``MB'' for the MD-AM model with Bayesian updates using the full conditionals from the joint model.  For both MB and MH, we specify the models according to \eqref{2a}--\eqref{2k}.

We presume that the population totals  $T_{X_1}=\sum_{i=1}^N x_{i1}$ and $T_{X_2}=\sum_{i=1}^N x_{i2}$ are known and use them as auxiliary margins for the hybrid missingness MD-AM models. To set $V_{X_1}$ and $V_{X_2}$ in each $\mathcal{S}$, we run a bespoke implementation of the ``mice'' package for one cycle, completing all missing values.  Using the completed data, we compute the expressions for the unbiased estimates of the variances of $\hat{T}_{X_1}$ and $\hat{T}_{X_2}$ under Poisson sampling, with unit nonrespondents' weights set as in  \eqref{known_weights_unit_res}.  
We also use the completed data from this preliminary run to estimate and set the initial values of all  model parameters. For MH and MB, in each $\mathcal{S}$ we execute 10,000 iterations of the sampler, discarding the first 5000 as burn-in.  From the remaining 5,000 posterior samples, we take every 100 posterior samples to generate $L = 50$ multiple imputation datasets. For MMH, we also create $L=50$ completed datasets using the default parameters of the ``mice'' package.  For both MMH and MH, we create donor pools for the hot deck based on $(X_1, X_2)$. 

We now give a sense of the steps for imputation of item nonresponse in MB and MH, as these steps are the most computationally intensive. Specifically, we present the conditionals for $X_2$ and $X_6$.  At any iteration $t$ of the estimation algorithm, we draw imputations  $x_{i2}^*$ for units with $(R_{i2}=1,U_i=0)$ from a Bernoulli distribution with probability
\begin{align}\label{eq:imputex1}
&P(X_2=x|R_2=1,U=0,\dots)=\frac{P(R_2=1|X_2=x,U=0,\dots)P(X_2=x|U=0,\dots)}{\sum_{x\in\{0,1\}}P(R_2=1|X_2=x,U=0,\dots)P(X_2=x|U=0,\dots)}\\
&\propto P(X_2=x|\dots)P(X_3|X_2=x,\dots) P(X_4|X_2=x,\dots)P(X_5|X_2=x,\dots)\\ \nonumber
&\times P(X_6|X_2=x,\dots) P(R_3|X_2=x,\dots)P(R_4|X_2=x,\dots)P(R_6|X_2=x,\dots).
\end{align}
Here, $P(R_2=1|X_1=x,U=0,\dots)$ cancels from \eqref{eq:imputex1} because of the ICIN assumption in \eqref{2h}, i.e., $R_2$ is conditionally independent of $X_2$. This is straightforward to compute, although  one must take care to avoid computational underflow.

For $X_6$, we use a rejection sampling step to impute values $x_{i6}^*$ for units with $(R_{i6}=1,U_i=0)$. Let $y$ represent the value of $x_{i6}^*$ from the current iteration of the sampler. We draw a potential replacement $y'$  from the normal distribution in \eqref{2g} using the current draw of the parameters  and completed data. 
Writing the normal density for $X_{6}$ evaluated at any $y$ as $f(y|\dots)$, we compute the ratio 
\begin{equation}
a= \frac{f(y'|\dots)P(R_{i2}|x_{i6}=y',\dots)P(R_{i3}|x_{i6}=y',\dots)P(R_{i4}|x_{i6}=y', \dots)}{f(y|\dots)P(R_{i2}|x_{i6}=y,\dots)P(R_{i3}|x_{i6}=y,\dots)P(R_{i4}|x_{i6}=y, \dots)}.\label{rejection}
\end{equation}
We randomly sample $u\sim \text{Unif}(0,1)$, and accept $y'$ as the new imputation when $u\leq a$ and keep $y$ as the imputation otherwise. With many missing values, it can be computationally expensive to get a sufficient effective sample size.  

One can reduce the computational burden resulting from \eqref{rejection} by presuming all $\phi_{j5}=0$ for $j\in\{2,3,4\}$ when estimating the model. In this case, the full conditional for $X_6$ need not include the probabilities for any $R_j$. Indeed, in general contexts, one can go further and assume a missing completely at random model for every $R_j$, which allows the analyst to ignore models for $\boldsymbol{R}$ entirely, although at the cost of a stronger assumption on missingness than ICIN.  In our simulations, we facilitate the computation for repeated sampling experiments by presuming $\phi_{j5}=0$ for $j\in\{2,3,4\}$ in both modeling and data generation, but we let other $\phi$ parameters take non-zero values in both modeling and data generation.  

After multiple imputation, we use the methods in \citet{rubin1987multiple} for inferences for the marginal totals for $(X_1, \dots, X_6)$ and several probabilities involving multiple categorical variables.  The full list of estimands, along with detailed results from the simulation runs, is in Appendix \ref{appendixB}.  For comparisons, we also specify a multiple imputation model that does not take advantage of $\mathcal{A}$. Specifically, we run the MH sampler to generate imputations for item nonresponse. However, we do not use the intercept matching algorithm for unit nonresponse. Instead, we set $\theta_1=\theta_2=0$ for estimation and, in each iteration, we sample $(X_1, X_2)$ for unit nonrespondents directly from \eqref{2b} and \eqref{2c} using the current draw of $(\Omega_1, \Omega_2)$.  We use the random hot deck to complete the unit nonrespondents' imputations. We refer to this method as IH, which stands for ICIN plus hot deck. For all imputation methods, we obtain inferences using Horvitz-Thompson point and variance estimators with weights from \eqref{known_weights_unit_res} and presume each completed dataset comes from a Poisson sampling design.

\begin{figure}[t]
\centering
\caption{Simulated relative RMSE and multiple imputation confidence interval coverage rates for population totals, two-way conditional probabilities, and joint probabilities for MB (fully model-based MD-AM), MH (MD-AM plus hot deck), MMH (MD-AM plus MICE plus hot deck), and IH (ICIN plus hot deck). Right panel compares the three MD-AM models, which have relatively similar performances.  Left panel compares IH to MMH, which tends to offer higher quality inferences than IH.
}
\includegraphics[width=\textwidth]{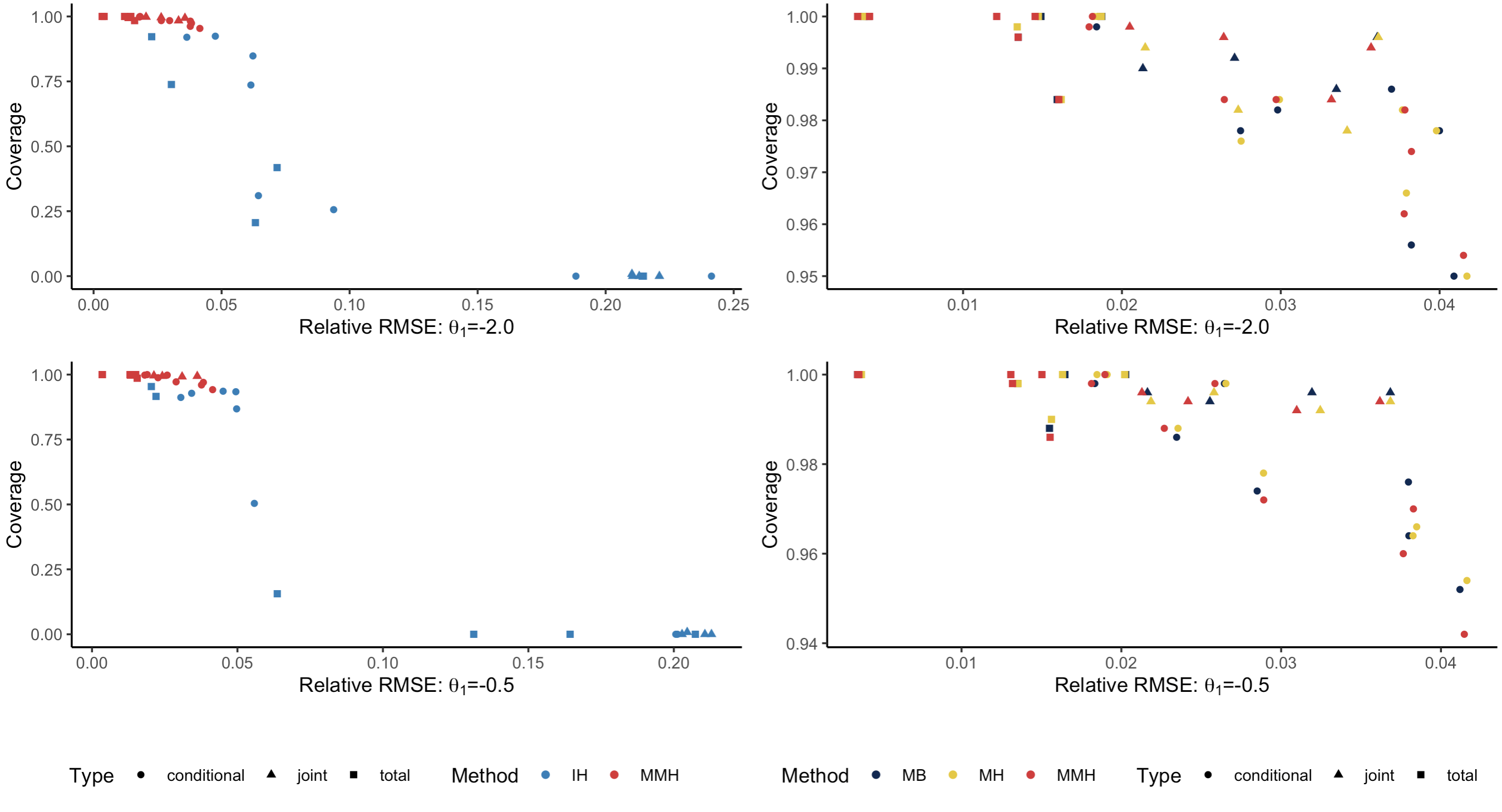}
\label{fig:sim2} 
\end{figure}

Figure \ref{fig:sim2} summarizes the results of the 500 simulation runs for both settings of $\theta_1$. Considering the right panel of the figure, the three hybrid missingness MD-AM models have similar repeated sampling properties with regard to relative RMSEs and empirical coverage rates. All of the relative RMSEs are less than 5\%, and all of the empirical coverage rates are close to or exceed the nominal 95\% level.   The key difference in the three models is computational running time. Using a standard laptop and programming in the software package R, one run of MB requires approximately 1.8 hours; one run of MH requires about 1.2 hours; and, one run of MMH takes about 25 seconds.  While undoubtedly the codes for MB and MH could be made more efficient, and 10,000 iterations may be more than necessary to generate $L=50$ independent draws for multiple imputation in this setting, the results suggest little analytical downsides from using MMH but potentially significant computational gains.  

Given the similar results of the three hybrid missingness MD-AM models, we compare the repeated sampling properties of MMH to those for IH.  The left panel of Figure \ref{fig:sim2} summarizes the results.  Unlike MMH, IH frequently offers low coverage rates and large relative RMSEs. These differences are most evident for $T_{X_1}$ and $T_{X_2}$; see Appendix \ref{appendixB} for the supporting numerical results. For these estimands, the multiple imputation point estimates for MMH closely match the known totals in $\mathcal{A}$ by design.  Thus, these point estimates are approximately unbiased, whereas those from IH suffer from bias due to the nonignorable nonresponse. Additionally, as shown in Appendix \ref{appendixB}, the repeated sampling variances of these point estimates for MMH are much smaller than the corresponding variances for IH, again due to the close matching to the known totals. The properties of MMH and IH are similar for estimands that are not affected as directly by the totals of $X_1$ and $X_2$, such as the conditional probabilities given $(X_1, X_2)$. As expected, the discrepancy in performance between  MMH and IH is more pronounced at the larger $|\theta_1|$, since the distributions of $X_1$ for unit respondents and nonrespondents are more dissimilar in that scenario. 

As evident in Appendix \ref{appendixB}, we also replicate a finding of \citet{akande2021leveraging} and \citet{tang2024using}, namely that for MD-AM models the average estimated variances from the rules of \citet{rubin1987multiple} can be larger than the corresponding (simulated) true variances, especially for the estimators that most directly utilize $\mathcal{A}$. In fact, for $T_{X_1}$ and $T_{X_2}$, the estimated multiple imputation variances for the MD-AM models are orders of magnitude larger than the corresponding true variances. This is because the variance formulas for multiple imputation  do not account for the known margins, making them a mismatch for the multiple imputation task and generally positively biased \citep{reiter:mime}. However, Appendix \ref{appendixB} shows that the average estimated variances from the rules of \citet{rubin1987multiple} for the MD-AM models are roughly of the same order as, although generally still larger than, the simulated true variances of the Horvitz-Thompson estimators computed with the complete data, i.e., before introduction of missing values. Thus, the widths of confidence intervals are reasonably similar to what would have been realized absent missing data. As noted by \citet{akande2021leveraging} and \citet{tang2024using}, research is needed to develop multiple imputation variance estimators that account for the known margins in this context. 

Finally, though not shown in the results reported here, the multiple imputation inferences using the weighted estimators 
are more accurate than inferences that treat the completed datasets like simple random samples and do not use the weights in inferences.

\section{Illustration with CPS Data}\label{sec:CPS}
\citet{tang2024using} apply the hybrid missingness MD-AM modeling framework using data from the 2018 Current Population Survey to analyze correlates of voter turnout. We illustrate our methodology using the same data with additional variables. Specifically, we use the intercept matching algorithm and hot deck imputation for unit nonresponse, and we use MICE imputation for item nonresponse. 

\subsection{Data}
The analysis data used by \citet{tang2024using} derive from the CPS November 2018 voter supplement for North Carolina. The data comprise $n=2926$ individuals, 913 of whom are unit nonrespondents. Whereas \cite{tang2024using} use five categorical variables, we consider the 11 variables listed in Table \ref{var:description}. We obtain these additional variables from the Integrated Public Use Microdata Series (IPUMS) website (\url{https://cps.ipums.org/cps/}). Among the variables in Table \ref{var:description}, we use the auxiliary marginal distributions for vote, sex, and race given in \citet{tang2024using}. In particular, the turnout rate for the voter-eligible population in North Carolina in 2018 was 49\% (\url{https://www.electproject.org/2018g}). As margins for sex and race, we use estimates from the 2018 ACS, which we treat as having negligible error at the state level. The percentages for sex are 52\% female and 48\% male. The percentages for race are 69.9\% white alone, 21.8\% black alone, 3.9\% Hispanic, and 4.4\% from other racial categories.

\begin{table}[t]
    \centering
    \caption{Description of variables in the CPS used in the voter turnout illustration. ``Difficulty'' is whether the person has any physical or cognitive difficulty. ``Duration'' is the length of time the person resides at the  current address. ``Proxy'' indicates whether the person completed the voter supplement themselves or whether a proxy provided information on their behalf. Variables presented in order of the visit sequence for the MICE algorithm for item nonresponse imputation.}
    \label{var:description}
    \begin{tabular}{lclr}
    \toprule
    \text{Variable} & Notation & Categories & Missing \%\\
    \toprule
    Difficulty & D &  0=No difficulty, 1=Has difficulty & 0\\
    Employment & T & 1=Employed, 2=Unemployed, 3=Not in labor force & 0\\
    Sex & S & 0=Male, 1=Female & 0.1 \\
    Race & E & 1=White, 2=Black, 3=Hispanic, 4=Other races & 2.0\\
    Marital & M & 1=Married, 2=Single, 3=Others & 2.5\\
    Education & C & 1=High school or less, 2=Some college, 3=Bachelor's and more & 4.0\\
    Age & A & Continuous, 18+ & 4.5\\
    Proxy & P & 0=Self, 1=Proxy & 14.5 \\
    Duration & B & 0=Less than 5 years, 1= At least 5 years & 16.0\\
    Vote & V & 0=Did not vote, 1=Voted & 18.0\\
    Family Income & F & 1=Less than 50k, 2=50k-100k, 3=100k+ & 27.0 \\
    \toprule
    \end{tabular}
\end{table}

\subsection{Models}\label{sec5.2}
To implement the MICE imputations for item nonresponse, we follow the advice in \citet{raghunathan2001multivariate} and order the variables from smallest to largest rates of missing values as shown in Table \ref{var:description}. We run a bespoke application of the ``mice'' package \citep{vanbuuren} using main effects for all variables and include item nonresponse indicators for all variables with missing items.  The default setting uses multinomial or logistic regressions for unordered categorical variables, proportional odds models for ordinal variables, and predictive mean matching for continuous variables. We cycle through the sequence of variables five times in each MICE run to create $L=50$ datasets completed for item nonresponse.

We use the intercept matching algorithm with hot deck imputation for unit nonresponse. This requires values of weights for the unit nonrespondents, which are not on the file. Unfortunately, we do not know the design weights for the participants either, as the weights on the IPUMS file have been adjusted for unit nonresponse. Using these adjusted weights, which we label as $w_i^a$, in \eqref{known_weights_unit_res} results in negative and nonsensical $w_i$ for analyses. Therefore, following \cite{tang2024using}, we modify \eqref{known_weights_unit_res} to generate analysis weights. For each individual $i\in \mathcal{S}$, we use an analysis weight  
\begin{equation}
w_i^* =  \left\{\begin{array}{lr}
        w_i^a (1-\frac{\sum_{j \in \mathcal{S}} U_j}{n}), & \text{if } U_i = 0\\
        \frac{\sum_{j \in \mathcal{S}} w_j^a}{n}, & \text{if } U_i = 1.
        \end{array}\right.
\label{approx_designW}
\end{equation}
This serves to approximate the design weights for unit respondents while ensuring that all $\sum_{i\in\mathcal{S}} w_i^*=\sum_{i\in\mathcal{S}} w_i^a$. We use $w_i^*$ in the intercept matching algorithm and for all survey-weighted analyses of the completed data. 

To set variances $V_{X_j}$ for use in the intercept matching algorithm, we use one of the datasets from the run of the ``mice'' software with imputations for item nonresponse. Using this completed data set, we compute the estimated variances of the estimated totals using the ``survey'' package in $R$ \citep{surveypackage}, assuming a probability proportional to size (pps) with replacement design using the adjusted survey weights on the file. Using these estimates, we set the standard deviation $\sqrt{V_S}= 87000$ for the total for Female. For race, we set $\sqrt{V_{E^2}}= 82000$ for the total for Black, $\sqrt{V_{E^3}}= 32000$ for the total for Hispanic, and $\sqrt{V_{E^4}}= 37000$ for the total for Other races. For vote, we set $\sqrt{V_V} = 85000$ for the total for Voted. 

For the variables in $\mathcal{A}$, to facilitate imputation of unit nonrespondents' data with the intercept matching algorithm, we use a pattern mixture model for $(U,S,E,V)$ specified as follows.
\begin{align}
    U_i&\sim\text{Bernoulli}(P(U_i=1)), \quad
 \text{logit}(\pi_{U})=\nu_0 \label{CPS: equ1}\\
    S_i&\sim\text{Bernoulli}(P(S_i=1)), \quad 
 \text{logit}(P(S_i=1))=\omega_{10}+\theta_{1} U_i\label{CPS: equ2}\\
    E_i&\sim\text{Multinomial}(P(E_i=e)), \quad \log\Bigg(\frac{P(E_i=e)}{P(E_i=1)}\Bigg)=\omega_{20}^e+\omega_{21}^e S_i+\theta_{2}^eU_i \label{CPS: equ3}\\
    V_i&\sim\text{Bernoulli}(P(V_i=1)), \label{CPS: equ4}\\
    &\quad 
    \text{logit}(P(V_i=1))=\omega_{30}+\omega_{31}S_i+\sum_{e=2}^4\omega_{32}^eI(E_i=e)+\sum_{e=2}^4\omega_{33}^eI(S_i=1,E_i=e)+\theta_3U_i.\nonumber
\end{align}

In each of the $L$ multiple imputation datasets completed for item nonresponse, we sample values of all $\Omega_j$ parameters from multivariate normal approximations. We obtain the means and covariances of these normal distributions by fitting maximum likelihood estimation routines in each completed dataset. We use the intercept matching algorithm for all $\theta_j$ parameters, which we then use to impute values of $(S, E, V)$ for the 913 unit nonrespondents.  Details are presented in Appendix \ref{appendixA}. For the remaining survey variables, we apply a random hot deck imputation. We create $2\times 2\times 4=16$ donor pools from the cross-classification of possible combinations of sex, race, and vote.  We randomly select a donor from the matched donor pool in the completed data and use its values as imputations of the remaining survey variables. 

With the $L=50$ completed datasets, we compute survey-weighted estimates of various population quantities using the ``survey'' package, assuming pps with replacement sampling and the weights in \eqref{approx_designW}. For comparison, we also use a bespoke implementation of MICE to impute missing values for item nonresponse only, disregarding the unit nonrespondents. We include the item nonresponse indicators as predictors in the conditional models.  We compute survey-weighted estimates using the 2013 survey participants and their adjusted weights $w_i^a$ from the IPUMS file. We refer to this as the ``MICE'' method.

\subsection{Analyses}
\begin{figure}[t]
\centering
\caption[Estimated proportions who voted in subgroups.]{Multiple imputation point estimates and 95\% confidence intervals for the proportion of voters in various groups using MMH (MD-AM plus MICE plus hot deck) with all 2926 surveyed individuals and MICE alone with adjusted weights using the 2013 participating individuals.}
\includegraphics[width=\textwidth]{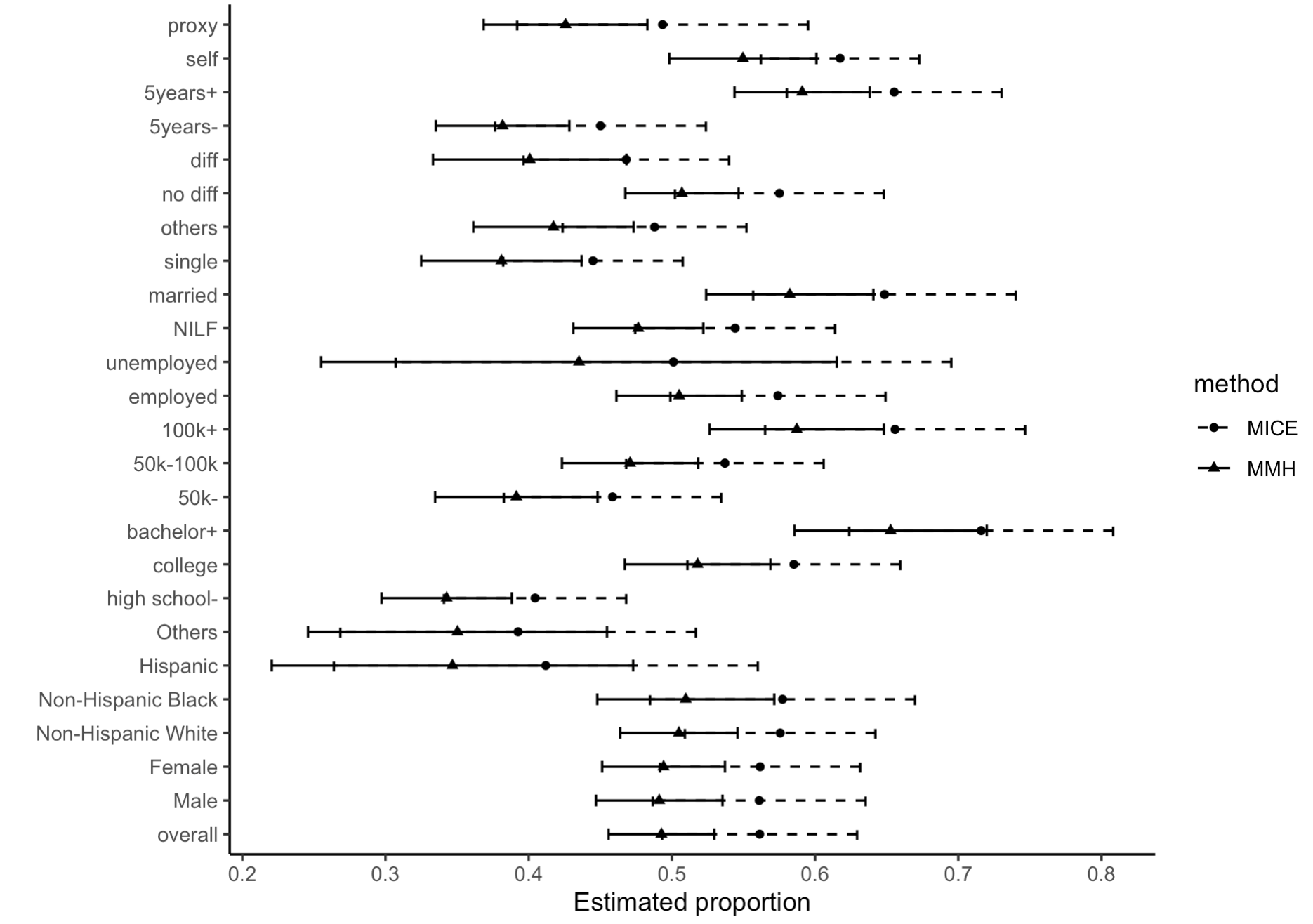}
\label{fig:apply-vag} 
\end{figure}

\begin{figure}[t]
\centering
\caption[Estimated proportions of voters in detailed subgroup of difficulty and education.]{Multiple imputation point estimates and 95\% confidence intervals for the proportion of voters in various subgroups based on the response to the questions on Difficulty and Education using MMH (MD-AM plus MICE plus hot deck) with all 2926 surveyed individuals and MICE alone with adjusted weights using the 2013 participating individuals.}
\includegraphics[width=\textwidth]{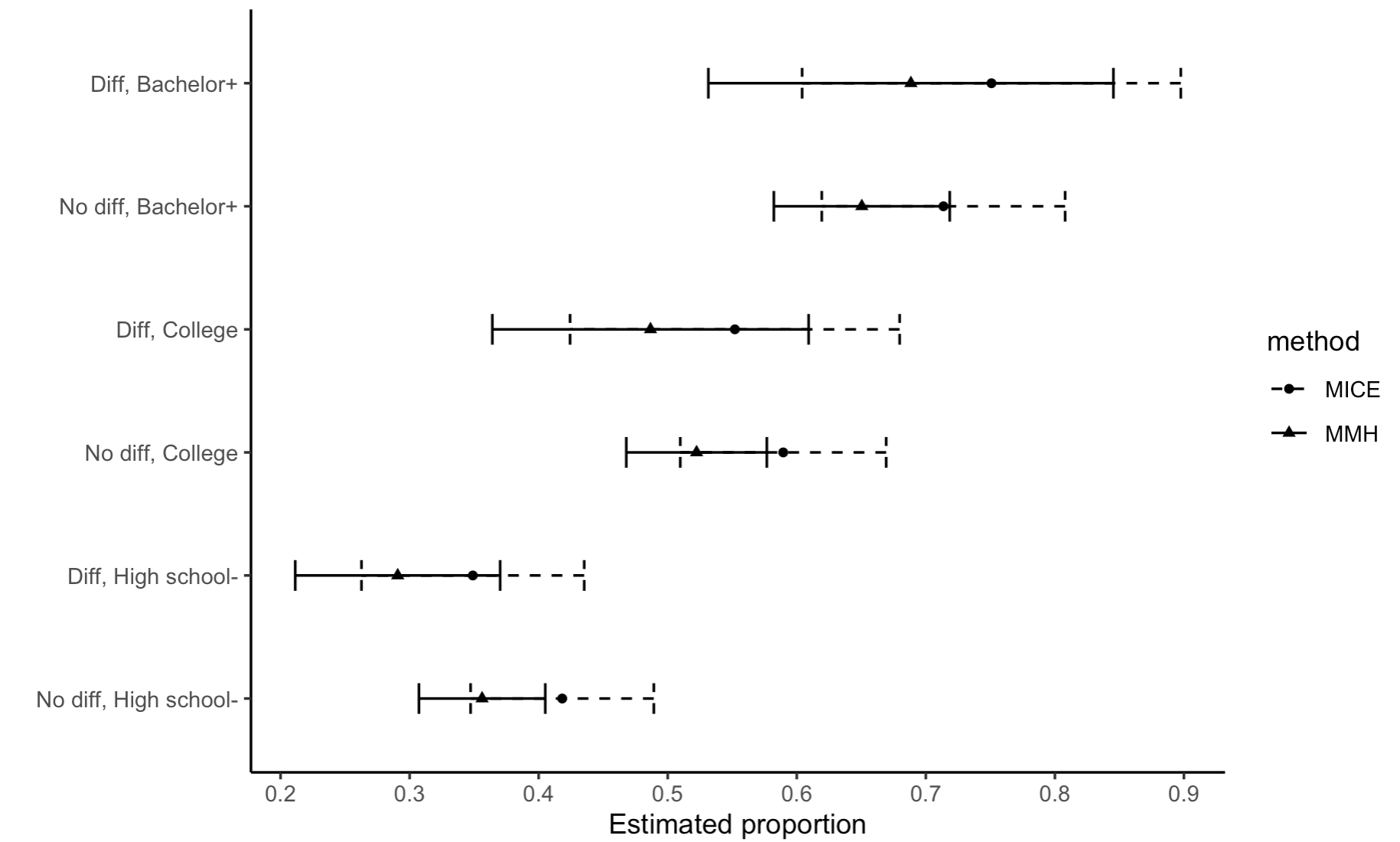}
\label{fig:apply-CD}
\end{figure}

Figure \ref{fig:apply-vag} displays the multiple imputation point estimates and 95\% confidence intervals for the proportion of voters in various demographic groups for both methods. The benefits of using the known margin for vote are apparent. For the overall turnout rate, the MMH estimate is 49.3\%, which is close to the true turnout of 49\%, whereas the estimate for MICE is too high at 56.1\%. Within groups, the turnout estimates from MMH are generally lower than the corresponding estimates from MICE. While we do not know the true turnout within groups, the rates from MMH seem more plausible than those from MICE. As examples, MMH estimates 58\% turnout among individuals who are married whereas MICE estimates 65\% turnout; MMH estimates 65\% turnout among individuals with a bachelor's degree or higher whereas MICE estimates 72\% turnout. The MICE turnout estimates appear implausibly high compared to the overall turnout rate of 49\%. On the other hand, both models estimate voters as older than nonvoters. For MMH, the average age is 52 (95\% CI: [50.9, 53.1]) for voters and 44 (95\% CI: [43.0, 46.0]) for nonvoters. For MICE, the average age is 52 (95\% CI: [50.7, 52.8]) for voters and 44 (95\% CI: [42.5, 45.5]) for nonvoters. 

Figure \ref{fig:apply-CD} displays the multiple imputation inferences for the proportion of voters in subgroups defined by cross-classifications of Difficulty and Education.  Using either imputation model, individuals who have physical or cognitive difficulties are estimated to have lower turnout than individuals who do not have these difficulties, at all levels of educational attainment.  The estimated turnout rates are lower with MMH than with MICE. For example, for people with difficulties who have at most high school degrees, the confidence interval for the turnout rate from MMH includes much lower values than the overall turnout of 49\%, whereas the MICE interval might lead one to conclude that these people have a turnout rate plausibly not too much lower than 49\%. We note that the relatively large standard errors result in overlapping confidence intervals for MMH and MICE.

\section{Discussion}\label{sec:conc}
The hybrid missingness MD-AM model presumes unit nonresponse potentially follows a missing not at random mechanism and item nonresponse follows an itemwise conditionally independent nonresponse mechanism. Allowing flexibility in the unit nonresponse mechansism is particularly appropriate when unit nonresponse is more concerning or prevalent than item nonresponse.  As noted by \citet{tang2024using}, developing computationally expedient MD-AM models for the case where survey item nonresponse is  missing not at random and survey unit nonresponse is missing at random is a topic for future research. 

An alternative paradigm for leveraging auxiliary margins is to adjust the design weights for survey participants, thereby discarding unit nonrespondents, and use imputation for item nonresponse. Using a fully multiple imputation approach has some potentially appealing features. The hybrid missingness MD-AM model can incorporate uncertainty in the marginal distributions themselves; for example, one can draw multiple plausible values of  $T_{X_j}$ from a normal distribution with mean and variance based on  published estimates, and use those plausible values in the runs of the intercept matching algorithm. Multiple imputation also could avoid potentially creating large weights due to the adjustments. Finally, the completed datasets may be preferable for analysts who wish to do analytical modeling without using survey weights.  However,  research is needed to understand the relative merits of each approach.

Finally, our approach relies on several choices that could be investigated and potentially improved.  First, we use random hot deck imputation, thereby disregarding donors' survey weights.  Future research could examine if there are any benefits to selecting donors within pools with unequal probabilities.  Second, the hybrid missingness MD-AM model presumes the survey data file does not include design weights for unit nonrespondents.  Future work could develop routines that facilitate the intercept matching algorithm when these design weights are available.  Third, we did not account for clustering or stratification features when creating completed records for unit nonrespondents.  Conceptually, if the relevant design information is available and released, the completed records could include such information. Examining the effectiveness of this approach, including the accuracy of variance estimates, is a topic for future work.

\section*{Acknowledgement}
The authors thank the editors for the invitation to contribute to the special anniversary issue of  {\em Survey Methodology}. The authors also thank Sunshine Hillygus for advice on the CPS data analysis.

\begin{appendices}
\section{Intercept Matching Algorithm for CPS Analysis}\label{appendixA}
In this section, we provide the steps for the intercept matching algorithm used in the CPS data analysis.  The steps presume the data have been completed for item nonresponse already. 
Following the notation in Section \ref{sec:CPS}, we impute missing values of $S$, $E$, and $V$ for unit nonrespondents by leveraging the known auxiliary margins for these variables. Let $n_U$ denote the number of unit nonrespondents.  The steps of the intercept matching algorithm proceed as follows.

\begin{enumerate}
\item Draw $\hat{T}_{S} \sim N(T_{S}, V_{S})$, where $T_S$ is the known total of Female sex and $V_S$ is set as described in Section \ref{sec5.2}.

\item Compute the number of times, denoted as $ n_{S}$, that the values of $\{S_{i}^*: U_i=1, i = 1, \dots, n\}$ should be assigned as 1. This is to ensure that the Horvitz-Thompson estimator for $T_S$ computed with the completed data approximates the sampled \( \hat{T}_{S} \) as closely as possible.  We have 
\[
n_{S} = \frac{\hat{T}_{S} - \sum_{i \in \mathcal{S}} w_{i}S_{i}^*I(U_i = 0)}{\sum_{i \in \mathcal{S}} w_{i} I(U_i = 1)/n_U}.
\]

\item Compute the maximum likelihood estimate of $\omega_{10}$, denoted as  $\hat{\omega}_{10}$, and its estimated variance $\hat{V}_{\hat{\omega}_{10}}$ based on \eqref{CPS: equ2} estimated using the unit respondents' data. Then, sample $\omega_{10}$ from its approximate posterior distribution, $ N(\hat{\omega}_{10}, \hat{V}_{\hat{\omega}_{10}})$. 

\item  Compute the proportion of unit nonrespondents that should be assigned $S_i^*= 1$, denoted as \( \hat{p}_{S} = n_{S} / n_U \). Calculate $\theta_1$ so that \( \text{logit}(\hat{p}_S) = \sum_{i:U_i=1}(\omega_{10} + \theta_1)/n_U \).

\item Draw imputations of $S_i^*$ for unit nonrespondents from
\[
S_i^*|U_i = 1 \sim \text{Bernoulli}\left( \frac{\exp(\omega_{10} + \theta_1)}{1 + \exp(\omega_{10} + \theta_1)} \right).
\]

\item Draw \( \hat{T}_{E^e} \sim N(T_{E^e}, V_{E^e}) \) for each level $e \in \{2,3,4\}$, where $T_{E^e}$ is the known total number of people with race at level $e$ and $V_{E^e}$ is set as described in Section \ref{sec5.2}.

\item Compute the number of imputed values \( n_{E^e} \) for each level $e \in \{2,3,4\}$ so that 
\[
n_{E^e} = \frac{\hat{T}_{E^e} - \sum_{i \in \mathcal{S}} w_{i}I(U_i = 0, E^*_i=e)}{\sum_{i \in \mathcal{S}} w_{i} I(U_i = 1)/n_U}. 
\]

\item Compute the maximum likelihood estimate of $\bm{\omega}_2=(\omega_{20}^2, \omega_{21}^2, \omega_{20}^3, \omega_{21}^3,\omega_{20}^4, \omega_{21}^4)^T$, denoted as $\hat{\bm{\omega}}_2$, and the corresponding estimated covariance matrix  \( \hat{\Sigma}_{\hat{\bm{\omega}}_2} \) based on \eqref{CPS: equ3} using the unit respondents' data. Then, sample $\bm{\omega}_2$ from its approximate posterior distribution, $N(\hat{\bm{\omega}}_2, \hat{\Sigma}_{\hat{\bm{\omega}}_2})$.

\item Compute the proportions \( \hat{p}_{E^e} = n_{E^e} / n_U \) for $e=2,3,4$ and $\hat{p}_{E^1}=1-\sum_{e=2}^4\hat{p}_{E^e}$ for base level $e=1$. Set $\theta_2^{e}$ for $e=2,3,4$ so that \[
\sum_{i:U_i=1}(\omega_{20}^{e}+\omega_{21}^{e}S^*_i+\theta_2^{e})/n_U  
= \text{log}(\hat{p}_{E^e}/\hat{p}_{E^1}).
\]

 \item Draw imputations of $E_i^*$ for unit nonrespondents using the probability mass function,
\[
P(E_i^*=e | U_i=1, S^*_i)=\frac{\exp{(\omega_{20}^{e}+\omega_{21}^{e}S_i^*+\theta_2^{e})}}{1+\sum_{e=2}^4 \exp{(\omega_{20}^{e}+\omega_{21}^{e}S_i^*+\theta_2^{e})}} \quad \text{for } e=2,3,4, 
\]
and \[
P(E_i^*=1)=1-\sum_{e=2}^4P(E_i^*=e).\]

\item Draw $\hat{T}_{V} \sim N(T_{V}, V_{V})$ where $T_V$ is the known total of Voted and $V_V$ is set as described in Section \ref{sec5.2}.

\item Compute the number of times, denoted as $ n_{V}$, that $\{V^*_{i}: U_i=1, i=1, \dots, n\}$ should be assigned as 1. We have
\[
n_{V} = \frac{\hat{T}_{V} - \sum_{i \in \mathcal{S}} w_{i}V_{i}^*I(U_i = 0)}{\sum_{i \in \mathcal{S}} w_{i} I(U_i = 1)/n_U}.
\]

\item Compute the maximum likelihood estimate of $\bm{\omega}_3=(\omega_{30},\omega_{31},\omega_{32}^2, \omega_{33}^2, \omega_{32}^3, \omega_{33}^3, \omega_{32}^4, \omega_{33}^4)^T$, denoted as  $\hat{\bm{\omega}}_3$, and its estimated covariance matrix $\hat{\Sigma}_{\hat{\bm{\omega}}_3}$ based on \eqref{CPS: equ4} using the unit respondents' data. Then, sample $\bm{\omega}_3$ from its approximate posterior distribution $ N(\hat{\bm{\omega}}_3, \hat{\Sigma}_{\hat{\bm{\omega}}_3})$. 

\item Compute the proportion of unit nonrespondents that should be assigned $V_i^*=1$, denoted as \( \hat{p}_{V} = n_{V} / n_U \). Calculate $\theta_3$ so that \[ 
\sum_{i:U_i=1}\Bigg(\omega_{30}+\omega_{31}S^*_i+\sum_{e=2}^4\omega_{32}^eI(E^*_i=e)+\sum_{e=2}^4\omega_{33}^eI(S^*_i=1,E^*_i=e)+\theta_3\Bigg)/n_U = \text{logit}(\hat{p}_V) \].

\item Draw imputations of $V_i^*$ for unit nonrespondents from
\begin{equation}
V_i^*|U_i = 1, S_i^*, E^*_i \sim \text{Bernoulli}(\pi_{iV})
\end{equation}
where
\begin{equation}
\pi_{iV} = \frac{\exp(\omega_{30}+\omega_{31}S_i^*+\sum_{e=2}^4\omega_{32}^eI(E_i^*=e)+\sum_{e=2}^4\omega_{33}^eI(S_i^*=1,E_i^*=e)+\theta_3)}{1 + \exp(\omega_{30}+\omega_{31}S_i^*+\sum_{e=2}^4\omega_{32}^eI(E_i^*=e)+\sum_{e=2}^4\omega_{33}^eI(S_i^*=1,E_i^*=e)+\theta_3)}.
\end{equation}

\end{enumerate}

\section{Supporting Results from the Simulation Studies}\label{appendixB}

In this section, we present supporting results from the simulation studies of Section \ref{sec:sims}. Parameters for the simulation are as follows.
\begin{itemize} 
\item $\nu_u = -1.2$.  
\item $\omega_{10} = (0.06, -0.0002)$ and $\theta_1 \in (-2, -0.5)$.
\item $(\omega_{20}, \omega_{21} = (0.2, 0.4)$ and $\theta_2 = -2$. 
\item $(\omega_{30}, \omega_{31}, \omega_{32})=(0.2, 0.3, 0.1)$.
\item $(\omega_{40}, \omega_{41}, \omega_{42}, \omega_{43})=(0.2, 0.4, 0.4, 0.1)$.
\item $(\omega_{50}, \omega_{51}, \omega_{52}, \omega_{53}, \omega_{54})=(0.4, 1.2, -0.9, 0.1, 0.2)$ and $\sigma_{x_5}=0.5.$ 
\item $(\omega_{60}, \omega_{61}, \omega_{62}, \omega_{63}, \omega_{64}, \omega_{65})=(0.4, 1.2, -0.9, 0.1, -0.1, 0.1)$, and $\sigma_{x_6}=0.5$.  
\item For $j \in \{2, 3, 4\}$, we set $(\phi_{j0}, \dots, \phi_{j5}) = (-1.4, 0.1, 0.1, 0.1, 0.1, 0)$. For $j=6$, we set $(\phi_{j0}, \dots, \phi_{j5}) = (-1.4, 0.1, 0.1, 0.1, 0.1, 0.1)$.
\end{itemize}

Table \ref{Result2:total} and Table \ref{Result2:prob} display results from the 500 repeated simulations runs for the three hybrid missingness MD-AM models and for the ICIN plus hot deck model described in Section \ref{sec:sims}. Results for population totals are in Table \ref{Result2:total}, and results for population percentages are in Table \ref{Result2:prob}. 

\begin{table}[t]
\centering
\caption[Results of population totals when data subject to unit and item nonresponse.]{Results of population totals when data subject to unit and item nonresponse under $\theta_1=-2.0$ and $\theta_1=-0.5$. Here, ``MB'' stands for fully model-based MD-AM, ``MH'' stands for MD-AM plus hot deck, ``MMH'' stands for MD-AM plus MICE plus hot deck, and ``IH'' stands for ICIN plus hot deck.   ``Pre'' stands for the variance of the Horvitz-Thompson estimator based on the complete data before introduction of missing values. Variances are on the scale of $10^8$.}
\label{Result2:total}
\resizebox{\textwidth}{!}{%
\begin{tabular}{lrrcrrrrrrrrrrcccc}
\toprule
& \multicolumn{4}{c}{Absolute Percent Bias (\%)} & \multicolumn{4}{c}{CI Coverage (\%)} & \multicolumn{5}{c}{Variance} & \multicolumn{4}{c}{Avg. Est. Var} \\
\cmidrule(lr){2-5} \cmidrule(lr){6-9} \cmidrule(lr){10-14} \cmidrule(lr){15-18}
& MB& MH & MMH & IH & MB & MH & MMH & IH & Pre & MB & MH & MMH & IH & MB & MH & MMH& IH \\
\midrule
\multicolumn{6}{l}{Scenario with $\bm{\theta_1=-2.0}$}\\
$\,\,\,T_{X_1}$ & .05 & .05 & .06 & 21.3 & 100.0 & 100.0 & 100.0 & 0.0 & 14.0 & 0.3 & 0.3 & 0.3 & 8.3 & 25.1 & 25.0 & 25.1 & 11.9 \\
$\,\,\,T_{X_2}$ & .03 & .05 & .04 & 21.4 & 100.0 & 100.0 & 100.0 & 0.0 & 15.4 & 0.4 & 0.4 & 0.3 & 9.2 & 28.7 & 29.0 & 27.1 & 14.5 \\
$\,\,\,T_{X_3}$ & .01 & .00 & .12 & 1.6 & 98.4 & 98.4 & 98.4 & 92.2 & 20.4 & 10.3 & 10.6 & 10.3 & 9.8 & 15.7 & 15.4 & 15.4 & 15.2 \\
$\,\,\,T_{X_4}$ & .09 & .09 & .06 & 2.7 & 99.6 & 99.8 & 99.6 & 73.8 & 21.5 & 8.6 & 8.6 & 8.7 & 8.5 & 16.2 & 16.0 & 15.8 & 15.3 \\
$\,\,\,T_{X_5}$ & .81 & .81 & .04 & 6.0 & 100.0 & 100.0 & 100.0 & 20.6 & 41.6 & 11.5 & 11.2 & 10.8 & 27.1 & 69.4 & 65.4 & 67.4 & 35.6 \\
$\,\,\,T_{X_6}$ & 1.10 & 1.10 & .00 & 6.8 & 100.0 & 100.0 & 100.0 & 41.8 & 36.7 & 12.7 & 12.6 & 11.5 & 27.5 & 92.3 & 88.5 & 77.3 & 53.2 \\
\addlinespace
\multicolumn{6}{l}{Scenario with $\bm{\theta_1=-0.5}$} \\
$\,\,\,T_{X_1}$ & .02 & .01 & .02 & 6.1 & 100.0 & 100.0 & 100.0 & 15.6 & 16.4 & 0.3 & 0.3 & 0.3 & 8.6 & 26.7 & 26.8 & 26.3 & 11.8 \\
$\,\,\,T_{X_2}$ & .10 & .05 & .08 & 20.7 & 100.0 & 100.0 & 100.0 & 0.0 & 15.3 & 0.3 & 0.4 & 0.3 & 8.0 & 28.9 & 29.4 & 27.6 & 14.5 \\
$\,\,\,T_{X_3}$ & .12 & .11 & .02 & 1.3 & 98.8 & 99.0 & 98.6 & 95.4  & 20.3 & 9.7 & 9.9 & 9.8 & 9.8 & 15.4 & 15.3 & 15.3 & 15.1 \\
$\,\,\,T_{X_4}$ & .17 & .17 & .01 & 1.8 & 99.8 & 99.8 & 99.8 & 91.6 & 21.7 & 8.7 & 8.8 & 8.5 & 8.6 & 16.0 & 15.9 & 15.8 & 15.3 \\
$\,\,\,T_{X_5}$ & .95 & .93 & .08 & 13.0 & 100.0 & 100.0 & 100.0 & 0.0 & 40.5 & 11.8 & 11.7 & 11.1 & 26.9 & 71.2 & 67.6 & 69.9 & 35.6 \\
$\,\,\,T_{X_6}$ & 1.30 & 1.30 & .16 & 16.2 & 100.0 & 100.0 & 100.0 & 0.0 & 36.3 & 11.2 & 11.3 & 10.3 & 29.4 & 94.7 & 91.6 & 80.3 & 53.2 \\
\bottomrule
\end{tabular}
}
\end{table}

\begin{table}[!htbp]
\centering
\caption[Results of probabilities when data subject to unit and item nonresponse.]{Results of probabilities when data subject to unit and item nonresponse under under $\theta_1=-2.0$ and $\theta_1=-0.5$. Here, ``MB'' stands for  fully model-based MD-AM, ``MH'' stands for MD-AM plus hot deck, ``MMH'' stands for MD-AM plus MICE plus hot deck, and ``IH'' stands for ICIN plus hot deck. ``Pre'' stands for the variance of the Horvitz-Thompson estimator based on the complete data before introduction of missing values. Variances are on the scale of $10^{-4}$.}
\label{Result2:prob}
\resizebox{\textwidth}{!}{%
\begin{tabular}{lcccccrrrrccccccccc}
\toprule
& \multicolumn{5}{c}{Estimate} & \multicolumn{4}{c}{CI Coverage (\%)} & \multicolumn{5}{c}{Variance} & \multicolumn{4}{c}{Avg. Est. Var} \\
\cmidrule(lr){2-6} \cmidrule(lr){7-10} \cmidrule(lr){11-15} \cmidrule(lr){16-19}
& Truth & MB & MH & MMH & IH & MB & MH & MMH & IH & Pre & MB & MH & MMH & IH & MB & MH & MMH & IH \\
\midrule
\multicolumn{6}{l}{Scenario with $\bm{\theta_1=-2.0}$}\\ 
$\,\,\,X_1=0|X_2=0$ & .681 & .683 & .683 & .680 & .553 & 100.0 & 100.0 & 99.8 & 0.0 & 1.2 & 1.5 & 1.5 & 1.5 & 1.8 & 4.8 & 4.8 & 4.6 & 2.1 \\
$\,\,\,X_1=0|X_2=1$ & .477 & .474 & .474 & .477 & .449 & 97.8 & 97.6 & 98.4 & 31.0 & 1.5 & 1.6 & 1.6 & 1.6 & 1.3 & 2.6 & 2.6 & 2.5 & 1.3 \\
$\,\,\,X_2=0|X_1=0$ & .593 & .595 & .595 & .593 & .450 & 100.0 & 100.0 & 100.0 & 0.0  & 1.2 & 1.2 & 1.2 & 1.2 & 1.7 & 4.4 & 4.5 & 4.2 & 1.8 \\
$\,\,\,X_2=0|X_1=1$ & .384 & .381 & .381 & .384 & .350 & 97.4 & 97.8 & 97.4 & 25.6 & 1.6 & 2.2 & 2.2 & 2.2 & 1.4 & 3.5 & 3.5 & 3.2 & 1.6 \\
$\,\,\,X_4=0|X_3=0$ & .372 & .372 & .372 & .371 &.353 & 98.2 & 98.2 & 98.2 & 84.8 & 1.5 & 1.9 & 2.0 & 2.0 & 1.8 & 2.6 & 2.6 & 2.6 & 2.5 \\
$\,\,\,X_4=0|X_3=1$ & .338 & .339 & .339 & .338 & .322 & 95.6 & 96.6 & 96.2  & 73.6 & 1.3 & 1.6 & 1.6 & 1.6 & 1.6 & 1.6 & 1.6 & 1.6 & 1.5 \\
$\,\,\,X_3=0|X_4=0$ & .427 & .427 & .427 & .426 & .417 & 95.4 & 95.0 & 95.4 & 92.4 & 2.1 & 3.1 & 3.2 & 3.1 & 3.0 & 3.1 & 3.1 & 3.1 & 3.2 \\
$\,\,\,X_3=0|X_4=1$ & .392 & .392 & .392 & .392 & .383 & 98.2 & 98.4  & 98.4 & 92.0 & 1.1 & 1.4 & 1.4 & 1.4 & 1.3 & 1.6 & 1.6 & 1.6 & 1.5 \\
\addlinespace
$\,\,\,X_2=0, X_3=0$ & .223 & .224 & .224 & .223 & .176 & 98.8 & 97.8 & 98.4 & 0.0 & .46 & .57 & .57 & .55 & .47 & .96 & .96 & .89 & .57 \\
$\,\,\,X_2=1, X_3=0$ & .181 & .180 & .180 & .181 & .218 & 99.0 & 99.6 & 99.4 & 0.8 & .43  & .43 & .42 & .42 & .66 & .77 & .77 & .72 & .70 \\
$\,\,\,X_2=0, X_3=1$ & .282 & .281 & .281 & .282 & .223 & 99.4 & 98.2 & 99.6 & 0.0 & .67 & .57 & .58 & .55 & .66 & 1.2 & 1.2 & 1.1 & .69 \\
$\,\,\,X_2=1, X_3=1$ & .314 & .315 & .315 & .314 & .382 & 99.2 & 99.4 & 99.8 & 0.0 & .65 & .44 & .44 & .41 & .92 & 1.3 & 1.3 & 1.2 & .93 \\
\addlinespace
$\,\,\,X_3=0|X_1=0,X_2=0$ & .450 & .451 & .451 &.448 & .451 & 96.2 & 96.6 & 96.0 & 98.2 & 2.2 & 3.2 & 3.2 & 3.2 & 3.7 & 3.1 & 3.1 & 2.9 & 4.5 \\
$\,\,\,X_3=0|X_1=0,X_2=1$ & .379 & .375 & .375 & .377 & .375 & 98.4 & 98.6 & 98.6 & 97.8 & 3.0 & 3.1 & 3.1 & 3.1 & 2.9 & 4.1 & 4.1 & 3.9 & 3.6 \\
$\,\,\,X_3=0|X_1=1,X_2=0$ & .425 & .430 & .430 & .426 & .430 & 97.6 & 97.2 & 97.2 & 97.4 & 4.6 & 4.4 & 4.4 & 4.4 & 4.0 & 6.1 & 6.1 & 5.7 & 5.5 \\
$\,\,\,X_3=0|X_1=1,X_2=1$ & .355 & .354 & .354 & .355 & .354 & 97.4  & 97.4 & 97.0 & 96.2 & 2.8 & 2.8 & 2.8 & 2.8 & 2.5 & 3.5 & 3.5 & 3.4 & 2.8 \\
$\,\,\,X_4=0|X_1=0,X_2=0$ & .437 & .439 & .439 &.436 & .439 & 98.0 & 98.0 & 96.2 & 97.8 & 2.1 & 2.7 & 2.7 & 2.7 & 3.1 & 3.1 & 3.1 & 2.8 & 4.5 \\
$\,\,\,X_4=0|X_1=0,X_2=1$ & .340 & .338 & .338 & .340 & .338 & 97.8  & 97.6 & 97.6 & 97.8 & 2.6 & 3.1 & 3.0 & 3.1 & 2.9 & 3.9 & 3.9 & 3.7 & 3.4 \\
$\,\,\,X_4=0|X_1=1,X_2=0$ & .342 & .346 & .346 & .342 & .346 & 98.6 & 98.8  & 98.4 & 98.2 & 4.0 & 3.7 & 3.7 & 3.7 & 3.5 & 5.5 & 5.5 & 5.2 & 5.0 \\
$\,\,\,X_4=0|X_1=1,X_2=1$ & .256 & .255 & .255 & .256 & .255 & 98.6  & 97.4 & 97.8 & 97.4 & 2.2 & 2.2 & 2.2 & 2.3 & 1.9 & 2.9 & 2.9 & 2.8 & 2.3 \\
\addlinespace
\multicolumn{6}{l}{Scenario with $\bm{\theta_1=-0.5}$}\\
$\,\,\,X_1=0|X_2=0$ & .581 & .583 & .583 & .582 & .552 & 100.0 & 100.0 & 100.0 & 50.4 & 1.4 & 1.2 & 1.2 & 1.2 & 2.0 & 5.3 & 5.3 & 5.0 & 2.1 \\
$\,\,\,X_1=0|X_2=1$ & .457 & .455 & .455 & .457 & .449 & 98.8 & 98.8 & 98.8 & 91.2 & 1.5 & 1.1 & 1.1 & 1.1 & 1.3 & 1.8 & 1.8 & 1.7 & 1.4 \\
$\,\,\,X_2=0|X_1=0$ & .562 & .563 & .563 & .561 & .449 & 100.0 & 100.0 & 99.8 & 0.0  & 1.4 & 1.1 & 1.1 & 1.0 & 1.6 & 3.5 & 3.5 & 3.3 & 1.8 \\
$\,\,\,X_2=0|X_1=1$ & .437 & .435 & .435 & .436 & .350 & 99.8 & 99.8 & 99.8 & 0.0 & 1.4 & 1.3 & 1.3 & 1.3 & 1.4 & 3.9 & 3.9 & 3.6 & 1.6 \\
$\,\,\,X_4=0|X_3=0$ & .365 & .366 & .366 & .365 &.354 & 97.2 & 96.6 & 97.0 & 93.4 & 1.5 & 1.9 & 2.0 & 2.0 & 1.9 & 2.5 & 2.5 & 2.5 & 2.5 \\
$\,\,\,X_4=0|X_3=1$ & .333 & .334 & .334 & .334 & .322 & 96.2 & 96.4 & 96.0  & 86.8 & 1.3 & 1.6 & 1.6 & 1.6 & 1.5 & 1.6 & 1.6 & 1.6 & 1.5 \\
$\,\,\,X_3=0|X_4=0$ & .425 & .426 & .426 & .425 & .418 & 94.8 & 95.4 & 94.2 & 93.6 & 2.2 & 3.1 & 3.1 & 3.1 & 3.1 & 3.1 & 3.1 & 3.1 & 3.2 \\
$\,\,\,X_3=0|X_4=1$ & .391 & .391 & .391 & .391 & .383 & 97.4 & 97.8  & 97.2 & 92.8 & 1.1 & 1.3 & 1.3 & 1.3 & 1.3 & 1.6 & 1.6 & 1.5 & 1.5 \\
\addlinespace
$\,\,\,X_2=0, X_3=0$ & .220 & .222 & .222 & .220 & .176 & 99.6 & 99.2 & 99.2 & 0.0 & .46 & .47 & .49 & .47 & .45 & .95 & .96 & .88 & .57 \\
$\,\,\,X_2=1, X_3=0$ & .182 & .181 & .181 & .182 & .219 & 99.6 & 99.4 & 99.4 & 0.8 & .44  & .44 & .44 & .43 & .69 & .76 & .78 & .73 & .70 \\
$\,\,\,X_2=0, X_3=1$ & .281 & .279 & .279 & .281 & .222 & 99.0 & 99.6 & 99.4 & 0.0 & .65 & .49& .49 & .46 & .55 & 1.2 & 1.2 & 1.1 & .68 \\
$\,\,\,X_2=1, X_3=1$ & .316 & .317 & .317 & .317 & .383 & 99.6 & 99.4 & 99.6 & 0.0 & .65 & .48 & .46 & .45 & .87 & 1.3 & 1.3 & 1.3 & .93 \\
\addlinespace
$\,\,\,X_3=0|X_1=0,X_2=0$ & .450 & .452 & .452 &.450 & .452 & 96.6 & 96.6 & 96.2 & 97.8 & 2.6 & 3.0 & 3.0 & 3.0 & 3.3 & 3.5 & 3.5 & 3.3 & 4.5 \\
$\,\,\,X_3=0|X_1=0,X_2=1$ & .379 & .375 & .376 & .378 & .376 & 97.8 & 97.6 & 97.8 & 97.0 & 3.1 & 3.3 & 3.4 & 3.3 & 3.0 & 4.2 & 4.2 & 4.0 & 3.6 \\
$\,\,\,X_3=0|X_1=1,X_2=0$ & .426 & .430 & .430 & .426 & .430 & 97.8 & 97.6 & 97.2 & 98.2 & 3.3 & 3.4 & 3.4 & 3.4 & 3.6 & 4.8 & 4.8 & 4.5 & 5.5 \\
$\,\,\,X_3=0|X_1=1,X_2=1$ & .355 & .353 & .353 & .355 & .354 & 97.2  & 97.2 & 97.0 & 96.2 & 2.6 & 2.8 & 2.7 & 2.8 & 2.5 & 3.3 & 3.3 & 3.2 & 2.8 \\
$\,\,\,X_4=0|X_1=0,X_2=0$ & .437 & .441 & .441 &.437 & .441 & 96.4 & 96.2 & 95.0 & 97.6 & 2.4 & 3.2 & 3.1 & 3.1 & 3.5 & 3.5 & 3.5 & 3.3 & 4.5 \\
$\,\,\,X_4=0|X_1=0,X_2=1$ & .340 & .338 & .338 & .340 & .338 & 97.4  & 97.6 & 97.2 & 96.6 & 2.7 & 3.2 & 3.1 & 3.2 & 2.8 & 4.0 & 4.0 & 3.8 & 3.4 \\
$\,\,\,X_4=0|X_1=1,X_2=0$ & .342 & .346 & .346 & .342 & .346 & 97.6 & 97.0  & 97.2 & 97.6 & 3.0 & 3.4 & 3.4 & 3.4 & 3.6 & 4.4 & 4.4 & 4.1 & 5.0 \\
$\,\,\,X_4=0|X_1=1,X_2=1$ & .256 & .255 & .255 & .256 & .255 & 98.2  & 98.4 & 97.6 & 97.4 & 2.1 & 2.1 & 2.1 & 2.1 & 1.9 & 2.8 & 2.8 & 2.7 & 2.3 \\
\bottomrule
\end{tabular}
}
\end{table}

\end{appendices}

\clearpage

\bibliographystyle{chicago}
\bibliography{ref.bib}

\end{document}